\documentclass[11pt]{article}

\usepackage[T1]{fontenc}
\usepackage[sc]{mathpazo}
\usepackage{amsmath}
\usepackage{amssymb}
\usepackage{enumerate}
\usepackage{amsthm}
\usepackage{amsfonts,mathrsfs}

\usepackage{hyperref}
\hypersetup{pdfpagemode=UseNone}

\newcommand{\setft}[1]{\mathrm{#1}}
\newcommand{\lin}[1]{\setft{L}\left(#1\right)}
\newcommand{\density}[1]{\setft{D}\left(#1\right)}
\newcommand{\unitary}[1]{\setft{U}\left(#1\right)}
\newcommand{\trans}[1]{\setft{T}\left(#1\right)}

\def\complex{\mathbb{C}}

\def\natural{\mathbb{N}}

\def\I{\mathbb{1}}

\newenvironment{mylist}[1]{\begin{list}{}{
    \setlength{\leftmargin}{#1}
    \setlength{\rightmargin}{0mm}
    \setlength{\labelsep}{2mm}
    \setlength{\labelwidth}{8mm}
    \setlength{\itemsep}{0mm}}}
    {\end{list}}

\def\ot{\otimes}

\newcommand{\iinner}[2]{\langle #1 | #2\rangle}
\newcommand{\out}[2]{| #1\rangle\langle #2 |}

\newcommand{\defeq}{\stackrel{\smash{\textnormal{\tiny def}}}{=}}

\newcommand{\Pa}[1]{\left(#1\right)}

\newcommand{\Br}[1]{\left[#1\right]}
\newcommand{\set}[1]{\{#1\}}

\newcommand{\ket}[1]{|#1\rangle}

\DeclareMathOperator{\trace}{Tr}

\newcommand{\Ptr}[2]{\trace_{#1}\Pa{#2}}

\newcommand{\Tr}[1]{\Ptr{}{#1}}

\def\cH{\mathcal{H}}
\def\cK{\mathcal{K}}

\def\rH{\mathrm{H}}
\def\rM{\mathrm{M}}
\def\rS{\mathrm{S}}

\newtheorem{thrm}{Theorem}[section]

\theoremstyle{definition}

\newtheorem{remark}[thrm]{Remark}
\newtheorem{exam}[thrm]{Example}
\numberwithin{equation}{section}

\newcounter{questionnumber}

\begin{document}

\title{\Large Comment on "Convergence of macrostates under reproducible processes"
[Phys. Lett. A 374: 3715-3717 (2010)]}

\author{Lin Zhang$^\text{a}$\footnote{E-mail: godyalin@163.com; linyz@zju.edu.cn; linyz@hdu.edu.cn}\ , Junde Wu$^\text{b}$\footnote{E-mail: wjd@zju.edu.cn}\ , Shao-Ming Fei$^\text{c}$\footnote{E-mail:
feishm@cnu.edu.cn}\\[1mm]
  {\small $^\text{a}$\it Institute of Mathematics, Hangzhou Dianzi University, Hangzhou 310018, PR~China}\\
  {\small $^\text{b}$\it Department of Mathematics, Zhejiang University, Hangzhou 310027, PR China}\\
  {\small $^\text{c}$\it School of Mathematics of Sciences, Capital Normal University, Beijing 100048, PR China}}
\date{}
\maketitle \mbox{}\hrule\mbox\\
\begin{abstract}

In this Letter, two counterexamples show that the superadditivity
inequality of relative entropy is not true even for the full-ranked
quantum states. Thus, an inequality of quantum channels and
complementary channels is not also true. Finally, a conjecture of
weak superadditivity inequality is presented.

\end{abstract}
\mbox{}\hrule\mbox\\

\section{Introduction}

Let ${\cH}$ and ${\cK}$ be two finite-dimensional \emph{Hilbert spaces}, $\lin{\cH, \cK}$ be the set of all linear operators from $\cH$ to $\cK$, if $\lin{\cH} = \lin{\cH}$, we denote $\lin{\cH, \cK}$ for $\lin{\cH}$,
$\trans{\cH, \cK}$ the set of all linear super-operators from $\lin{\cH}$ to $\lin{\cK}$. A super-operator
$\Lambda\in \trans{\cH, \cK}$ is said to be a \emph{completely positive
linear map} if for each $k \in \natural$,
$$
\Lambda\ot \I_{\rM_{k}(\complex)}: \lin{\cH} \ot \rM_{k}(\complex)
\to \lin{\cK}\ot \rM_{k}(\complex)
$$
is positive, where $\rM_{k}(\complex)$ denotes the set of all
$k\times k$ complex matrices. It follows from Choi's theorem
\cite{Choi} that every completely positive linear map $\Lambda$ has
a Kraus representation
$$
\Lambda = \sum_{\mu}\mathrm{Ad}_{M_{\mu}},
$$
that is, for every $X\in \lin{\cH}$,  $\Lambda (X) = \sum_{\mu}M_\mu
XM_\mu^\dagger$, where $\set{M_\mu}\subseteq \lin{\cH, \cK}$,
$\sum^K_{\mu=1}M^\dagger_\mu M_\mu= \I_{\cH}$, $M_\mu^\dagger$ is
the adjoint operator of $M_\mu$. A \emph{quantum channel} is just a
trace-preserving completely positive linear super-operator.

Let $\density{\cH}$ denote the set of all the density matrices
$\rho$ on $\cH$. The \emph{von Neumann entropy} $\rS(\rho)$ of
$\rho$ is defined by
$$
\rS(\rho) \defeq - \Tr{\rho\log\rho}.
$$
The \emph{relative entropy} of two mixed states $\rho$ and $\sigma$
is defined by
$$
\rS(\rho||\sigma) \defeq \left\{\begin{array}{ll}
                             \Tr{\rho(\log\rho -
\log\sigma)}, & \text{if}\ \mathrm{supp}(\rho) \subseteq
\mathrm{supp}(\sigma), \\
                             +\infty, & \text{otherwise}.
                           \end{array}
\right.
$$

The relative entropy is an very important quantity in quantum
information theory \cite{Ohya}. It satisfies many significant
relations such as monotonicity property under quantum channels
\cite{Lindblda}. In \cite{Petz1,Petz2}, Petz studied the strong
superadditivity of relative entropy. In \cite{Rau}, Rau derived a
monotonicity property of relative entropy under a reproducible
process. From which he obtained the following \emph{superadditivity
inequality} of relative entropy:
\begin{eqnarray}\label{eq:superadditivity}
\rS(\rho_{AB}||\sigma_{AB})\geqslant \rS(\rho_A||\sigma_A) +
\rS(\rho_B||\sigma_B),
\end{eqnarray}
where $\rho_{AB}$ and $\sigma_{AB}$ are macrostates on tensor space
$\cH_A\ot\cH_B$, $\rho_A$, $\rho_B$, $\sigma_A$ and $\sigma_B$ are
the reduced states of $\rho_{AB}$ and $\sigma_{AB}$,
respectively. Note that the inequality ~\eqref{eq:superadditivity} holds if
$\sigma_{AB}$ is a product state.

In this Letter, however, we show that the inequality ~\eqref{eq:superadditivity} is not true even for the full-ranked quantum states. Thus, an inequality of quantum channels and complementary channels is not also true. Finally, we present a conjecture of weak superadditivity inequality of relative entropy.

\section{Counterexamples}\label{sec2}

Firstly, we show that the superadditivity inequality ~\eqref{eq:superadditivity} is not true.

\begin{exam}
Let $\ket{\psi_X},\ket{\phi_X}\in\cH_X$ such that
$\iinner{\psi_X}{\phi_X}=0$, where $X=A,B$. Set
$$
\rho_{AB}=\out{\psi_A}{\psi_A}\ot\out{\psi_B}{\psi_B}
$$
and
$$
\sigma_{AB}= \lambda\out{\psi_A}{\psi_A}\ot\out{\psi_B}{\psi_B} +
(1-\lambda)\out{\phi_A}{\phi_A}\ot\out{\phi_B}{\phi_B},
$$
where $\lambda\in(0,1)$. We have
$$
\rS(\rho_{AB}||\sigma_{AB}) = \rS(\rho_A||\sigma_A) =
\rS(\rho_B||\sigma_B)= -\log(\lambda)>0,
$$
which implies that
$$
\rS(\rho_{AB}||\sigma_{AB}) < \rS(\rho_A||\sigma_A) +
\rS(\rho_B||\sigma_B).
$$
Thus, the inequality ~\eqref{eq:superadditivity} is violated.
\end{exam}

The following numerical example of the diagonal and
\emph{full-ranked} states $\rho_{AB}$ and $\sigma_{AB}$ given by M.
Mosonyi show that the inequality ~\eqref{eq:superadditivity} is also
not true.

\begin{exam}[Random research]\label{exam} Let
$$
\rho_{AB} = 0.1568\out{00}{00} + 0.7270\out{10}{10} + 0.0804\out{01}{01} + 0.0358\out{11}{11}
$$
and
$$
\sigma_{AB} = 0.3061\out{00}{00} + 0.4243\out{10}{10} + 0.1713\out{01}{01} + 0.0983\out{11}{11}.
$$
Thus we have
\begin{eqnarray}
\begin{cases}
\rho_A = 0.2372\out{0}{0} + 0.7628\out{1}{1}\\
\sigma_A = 0.4774\out{0}{0} + 0.5226\out{1}{1}
\end{cases}
\text{and~~}
\begin{cases}
\rho_B = 0.8838\out{0}{0} + 0.1162\out{1}{1}\\
\sigma_B = 0.7304\out{0}{0} + 0.2696\out{1}{1}.
\end{cases}
\end{eqnarray}
Apparently, all states here are invertible and
$$
\rS(\rho_{AB}||\sigma_{AB}) < \rS(\rho_A||\sigma_A) + \rS(\rho_B||\sigma_B)
$$
which contradicts with the superadditivity inequality again.
\end{exam}

\begin{remark}
Now, we show that an inequality of quantum channels and complementary channels is not also true since the superadditivity inequality is not hold.

In fact, Let
$\rho,\sigma\in\density{\cH}$. Let $\Phi$ be a quantum channel from
$\cH$ to $\cK$,
$$
\Phi = \sum^K_{\mu=1} \mathrm{Ad}_{M_\mu},
$$
where $M_\mu\in\lin{\cH,\cK}$ are Kraus operators such that
$\sum^K_{\mu=1}M^\dagger_\mu M_\mu= \I_{\cH}$. Let $\cH_E =
\complex^K$ be a complex Hilbert space with orthonormal basis
$\set{\ket{\mu}:\mu = 1,\ldots,K}$. Define
$$
V\ket{\psi}\defeq \sum_\mu M_\mu\ket{\psi}\ot \ket{\mu},\quad
\forall \ket{\psi}\in\cH.
$$
According to the Stinespring representation of quantum channels, one
has
$$
\Phi(\rho) = \Ptr{E}{V\rho V^\dagger}.
$$
The corresponding \emph{complementary channel} is given by
$$
\widehat{\Phi}(\rho) =\Ptr{\cK}{V\rho V^\dagger} =
\sum_{\mu,\nu=1}^K \Tr{M_\mu \rho M^\dagger_\nu}\out{\mu}{\nu}.
$$

That $V\in\lin{\cH,\cK\ot\cH_E}$ is a linear isometry, and for all $\tau\in\density{\cH}$,
$V\tau V^\dagger$ has, up to multiplicities of zero, the same
eigenvalues as $\tau$ are clear. Thus,
\begin{eqnarray}\label{eq:mono-channel}\nonumber
\rS(\rho||\sigma) &=& \rS(V\rho V^\dagger||V\sigma V^\dagger)\\\nonumber
&\geqslant& \rS\Pa{\Ptr{E}{V\rho
V^\dagger}||\Ptr{E}{V\sigma V^\dagger}} + \rS\Pa{\Ptr{\cK}{V\rho
V^\dagger}||\Ptr{\cK}{V\sigma
V^\dagger}}\\
&=& \rS(\Phi(\rho)||\Phi(\sigma)) +
\rS(\widehat{\Phi}(\rho)||\widehat{\Phi}(\sigma)).
\end{eqnarray}

Taking $\rho = \rho_{AB}$ and $\Phi(\rho_{AB})=\Ptr{B}{\rho_{AB}}$,
we have $\widehat{\Phi}(\rho_{AB}) = W\rho_B W^\dagger$ for some
linear isometry $W\in\lin{\cH_B,\cH_E}$. It follows from
inequality~\eqref{eq:mono-channel} that
\begin{eqnarray*}
\rS(\rho_{AB}||\sigma_{AB}) &\geqslant&
\rS(\Phi(\rho_{AB})||\Phi(\sigma_{AB})) +
\rS(\widehat{\Phi}(\rho_{AB})||\widehat{\Phi}(\sigma_{AB}))\\
&=& \rS(\rho_A||\sigma_A)
+ \rS(W\rho_B W^\dagger||W\sigma_B W^\dagger)\\
&=& \rS(\rho_A||\sigma_A) + \rS(\rho_B||\sigma_B),
\end{eqnarray*}
which coincides with inequality~\eqref{eq:superadditivity}. As \eqref{eq:superadditivity} is not true, inequality~\eqref{eq:mono-channel} is also not true. That is,
$$
\rS(\rho||\sigma) \ngeqslant \rS(\Phi(\rho)||\Phi(\sigma)) +
\rS(\widehat{\Phi}(\rho)||\widehat{\Phi}(\sigma)).
$$

In \cite{Winter}, Li and Winter proposed the following
question: For given quantum channel $\Phi$ from $\cH_A$ to $\cH_B$ and
quantum states $\rho,\sigma\in\density{\cH_A}$, does there exist a quantum
channel $\Psi$ from $\cH_B$ to $\cH_A$ with
$\Psi\circ\Phi(\sigma)=\sigma$ and
\begin{eqnarray}\label{eq:lower-bound}
\rS(\rho||\sigma) \geqslant \rS(\Phi(\rho)||\Phi(\sigma)) +
\rS(\rho||\Psi\circ\Phi(\rho)) ?
\end{eqnarray}
They answered this question affirmatively in the classical
case. However, the quantum case is still open. In view of this, we can ask the following questions:
\begin{enumerate}[(i)]
\item Can we have $\widehat{\Phi}(\rho)=\widehat{\Phi}(\sigma)$ if
$\rS(\rho||\sigma) = \rS(\Phi(\rho)||\Phi(\sigma))$ ?
\item What can be derived from
$\widehat{\Phi}(\rho)=\widehat{\Phi}(\sigma)$ ?
\item What can be derived from
$\rS(\rho||\sigma) =
\rS(\widehat{\Phi}(\rho)||\widehat{\Phi}(\sigma))$ ?
\end{enumerate}

For (i), M. Hayashi answered negatively in \cite{Hayashi}.

Let $\rho,\sigma$ and $\Phi$ be as follows:
$$
\rho = \sum_j \lambda_j(\rho)E_j,\quad \sigma = \sum_j
\lambda_j(\sigma)E_j,\quad \Phi(X) = \sum_j E_jXE_j,
$$
where $E_j$ is a projector operator and $\sum_j E_j =\I$. Then $\rS(\rho||\sigma) =
\rS(\Phi(\rho)||\Phi(\sigma))$ and
$$
\widehat\Phi(\rho) = \sum_j
\lambda_j(\rho)\out{j}{j},\quad\widehat\Phi(\sigma) = \sum_j
\lambda_j(\sigma)\out{j}{j}.
$$
It is clear that if $\lambda(\rho) \neq \lambda(\sigma)$, then
$\widehat\Phi(\rho)\neq \widehat\Phi(\sigma)$.

This showed that no matter how close together $\rho$ and $\sigma$
are,  the inequality~\eqref{eq:mono-channel} does not hold.
Therefore, it seems that the inequality~\eqref{eq:superadditivity}
does not hold even if $\rho_{AB}$ and $\sigma_{AB}$ are closer in
some sense.

\end{remark}

\section{Discussions}

It is said in \cite{Rau} that the second law of thermodynamics, i.e.,
any reproducible process increases entropy, similarly implies that under a reproducible
process macrostates become less distinguishable from the uniform distribution
\begin{eqnarray}
\rS(\mu_f||\I/\Tr{\I})\leqslant \rS(\mu_g||\I/\Tr{\I}),
\end{eqnarray}
where $\mu_f,\mu_g$ are the so-called \emph{generalized canonical
distribution}
$$
\mu_g \propto \exp\Br{\sum_a \lambda^a G_a},
$$
$\set{G_a}$ are the observables whose expectation values characterize
the system's macrostate. With properly adjusted Lagrange
parameters $\set{\lambda^a}$ this canonical state encodes information about
the relevant expectation values $\set{g_a \equiv \langle G_a\rangle_\mu}$, while discarding
(by maximizing entropy) all other information. The initial macrostate $\mu_g$ evolves under the
same reproducible process to the final macrostate $\mu_f$.

It follows that full-ranked states are some kind of macrostates and
a reproducible process (coarse-graining) could be a process which
maps a full-ranked state into another full-ranked state. As the
second law of thermodynamics reflects the fact that the macrostates
tend to be closer to equidistribution, Rau intuitively thought that
not only the distinguishability between \emph{any} macrostate and
the uniform distribution diminishes, but also the mutual
distinguishability (described by the relative entropy) between
\emph{arbitrary pairs} of macrostates decreases. Thus, he proposed
the following monotonicity inequality: for \emph{any} two initial
macrostates $\mu_g$ and $\mu_{g'}$ evolving under the same
reproducible process to final macrostates $\mu_f$ and $\mu_{f'}$,
respectively, the relative entropy will decrease:
\begin{eqnarray}\label{eq:mutul-dis}
\rS(\mu_f||\mu_{f'})\leqslant \rS(\mu_g||\mu_{g'}).
\end{eqnarray}
He concluded that the inequality ~\eqref{eq:mutul-dis} follows immediately if only one can show the monotonicity
relation
\begin{eqnarray}\label{eq:nonlinear-coarse-graining-yes}
\rS(\mu_{f(\rho)}||\mu_{f(\sigma)}) \leqslant
\rS(\rho||\sigma),
\end{eqnarray}
where $\mu_{f(\rho)},\mu_{f(\sigma)}$ are the final macrostates
evolved from $\rho,\sigma$ under the same reproducible process,
respectively.

The inequality~\eqref{eq:nonlinear-coarse-graining-yes} is the main
result of Rau in \cite{Rau}. When the removing correlations are
considered, Rau obtained the superadditivity
inequality~\eqref{eq:superadditivity} from
inequality~\eqref{eq:nonlinear-coarse-graining-yes}. Since the
inequality~\eqref{eq:superadditivity} is not true, so, the
inequality~\eqref{eq:nonlinear-coarse-graining-yes} is not also
true. Thus, one needs to reconsider some related results which based
on inequalities~\eqref{eq:superadditivity} and
\eqref{eq:nonlinear-coarse-graining-yes}, for instance, the Lemma B5
in \cite{Genoni}, etc.

\section{Weak superadditivity inequality: A conjecture}

Although inequalities \eqref{eq:superadditivity} and
\eqref{eq:mono-channel} are not valid in general, it seems that a
modified version of the results is possible. If $\rho$ equals to the
coarse-grained $\mu_f(\sigma)$, e.g., for the case of removing
correlations and $\rho_{AB} = \sigma_A\ot \sigma_B$, then $\gamma =
0$. $\gamma$ is bounded in the range [0, 1] and varies continuously
as a function of $\rho$. For $\rho$ within some finite neighborhood
of $\mu_f(\sigma)$ (for the case of removing correlations and
$\rho_{AB}$ within some finite neighborhood of $\sigma_A \ot
\sigma_B$), $\gamma$ should still remain strictly smaller than one.
Hence, while the strong monotonicity of the relative entropy may no
longer hold globally for arbitrary pairs of states, it may still
hold locally for nearby states within some finite region. Indeed,
pursuing an alternative approach (within the framework of
nonequilibrium thermodynamics) to prove exactly such local
convergence of macrostates is deserved.

In
\cite{Lin},
we have obtained the following result:

\emph{For given two quantum states $\rho,\sigma\in\density{\cH_d}$, one has
\begin{eqnarray}\label{eq:zhang}
\begin{cases}
\min_{U\in\unitary{\cH_d}} \rS(U\rho U^\dagger||\sigma) =
\rH(\lambda^{\downarrow}(\rho)||\lambda^{\downarrow}(\sigma)),\\
\max_{U\in\unitary{\cH_d}} \rS(U\rho U^\dagger||\sigma) =
\rH(\lambda^{\downarrow}(\rho)||\lambda^{\uparrow}(\sigma)),
\end{cases}
\end{eqnarray}
where $\sigma$ is full-ranked, $\lambda^{\downarrow}(\sigma)$ (resp.
$\lambda^{\uparrow}(\sigma)$) stands for the vector with all
eigenvalues of $\sigma$ arranged in decreasing (resp. increasing)
order, $\unitary{\cH_d}$ denotes the set of all unitary operators on
$\cH_d$; $\rH(p||q):=\sum_jp_j(\log p_j - \log q_j)$ is Shannon
entropy between two probability distribution $p=\set{p_j}$ and
$q=\set{q_j}$.}

Based on the above result, we propose the following
\emph{conjecture}: There exist unitary operators
$U_A\in\unitary{\cH_A}$, $U_B\in\unitary{\cH_B}$ and
$U_{AB}\in\unitary{\cH_A\ot\cH_B}$ such that
\begin{eqnarray}\label{eq:weaker-version}
\rS(U_{AB}\rho_{AB}U^\dagger_{AB}||\sigma_{AB})\geqslant
\rS(U_A\rho_AU^\dagger_A||\sigma_A) +
\rS(U_B\rho_BU^\dagger_B||\sigma_B),
\end{eqnarray}
where the reference state $\sigma_{AB}$ is required to be
full-ranked state. Indeed, our numerical calculations show that
inequality~\eqref{eq:weaker-version} is true. More specifically, the
relative entropy is \emph{weak superadditivity} in the following
sense
\cite[arXiv:1305.2023]{Lin}:
\begin{eqnarray}
\rH(\lambda^\downarrow(\rho_{AB})||\lambda^\uparrow(\sigma_{AB}))\geqslant
\rH(\lambda^{\downarrow}(\rho_A)||\lambda^{\downarrow}(\sigma_A)) +
\rH(\lambda^{\downarrow}(\rho_B)||\lambda^{\downarrow}(\sigma_B)),
\end{eqnarray}
where $\sigma_{AB}$ is a full-ranked state.

\subsection*{Acknowledgement}
We thank F. Brand\~{a}o, M. Hayashi, M. Mosonyi, M. Piani, J. Rau
and A. Winter for valuable comments. The first-named author would
like to thank Shunlong Luo for his proposal to optimization
in~\eqref{eq:zhang}, and  L.Z. is grateful for funding from Hangzhou
Dianzi University (KYS075612038). The work is also supported by
Natural Science Foundations of China (11171301, 10771191, 10471124
and 11275131) and the Doctoral Programs Foundation of Ministry of
Education of China (J20130061).


\end{document}